\documentclass[12pt]{article}
\usepackage{amsmath, amssymb, latexsym, biblatex}
\usepackage{hyperref}
\hypersetup{
colorlinks=true,
linkcolor=red,
urlcolor=black,
citecolor=blue
}
\begin{document}

\title{On the Absoluteness of Rotation.}
\date{}
\maketitle

\vspace{-15mm}

\begin{center}
\author{P\'{e}ter Hrask\'{o}}\\
Department of Theoretical Physics, University of P\'{e}cs, Hungary\\
\medskip
\author{D\'{a}vid Szepessy}\\
Student of the Eötvös Loránd University (ELTE), Budapest, Hungary
\end{center}
\bigskip

\begin{abstract}
We argue that in the general relativistic calculation of planetary orbits, the choice of a reference frame which is an obligatory condition in the Newtonian approach is replaced by an appropriate boundary condition on the solution of Einstein equation. Implications of this observation on the nature of rotation and the physical interpretation of the metric tensor are discussed.
\end{abstract}

In {\em Principia} [1] we read the following lines about Newton's famous bucket experiment:

\begin{quotation}
If a vessel, hung by a long cord, is so often turned about that the cord is strongly twisted, then filled with water, and held at rest together with the water; after, by the sudden action of another force, it is whirled about in the contrary way, and while the cord is untwisting itself, the vessel continues for some time this motion; the surface of the water will at first be plain, as before the vessel began to move; but the vessel by gradually communicating its motion to the water, will make it begin sensibly to revolve, and recede by little and little, and ascend to the sides of the vessel, forming itself into a concave figure... This ascent of the water shows its endeavour to recede from the axis of its motion; and the true and absolute circular motion of the water, which is here directly contrary to the relative, discovers itself, and may be measured by this endeavour. ... And therefore, this endeavour does not depend upon any translation of the water in respect to ambient bodies, nor can true circular motion be defined by such translation. ...; but relative motions...are altogether destitute of any real effect. ... 
\end{quotation}

In the 19th century, this experiment could have been rephrased in the following way: When calculating planetary orbits, only the law of universal gravitation is taken into account on the right-hand side of Newton's equation, no inertial forces are included. With this, we automatically assume that we are in an inertial (specifically non-rotating) frame of reference. However, such an inertial system of the size of the Solar System cannot be realized, so instead, we check the correctness of the calculations in relation to distant stars and find that the observations confirm the calculations very accurately. Why?

For the sake of brevity, we will refer to the question formulated in the original bucket experiment and in its form outlined in the previous paragraph as {\em Newton's problem}.

Newton’s problem is logically similar to the problem of the equality of the inertial and gravitational mass (which, of course, is also a "Newton’s problem"). Knowing the genesis of the general theory of relativity, however, we are inclined to consider the latter to be much more fundamental than the former. But it is by no means certain that this is no more than hindsight. Perhaps it is enough to refer to Lorentz's electromagnetic theory of gravity [2], which he published in 1900.  According to the priorities in physics of those times, Lorentz considered the elimination of the action at a distance from Newton's theory to be much more important than explaining the equality of inertial and gravitational mass. His ambitious theory built on the analogy of electrodynamics transformed gravitation into retarded interaction between celestial bodies, but the equality of the two kinds of mass in his theory remained a mystery to be solved in the future. We believe that within the framework of Newtonian dynamics, Newton’s problem is as important as the equality of the inertial and gravitational masses. Below, however, we argue that the general theory of relativity solves the former problem too in a natural way.

Newton's answer to Newton's problem was that rotation is absolute [1]:
\begin{quotation} 
It is indeed a matter of great difficulty to discover, and effectually to distinguish, the true motions of particular bodies from the apparent; because the parts of that immovable space in which these motions are performed, do by no means come under the observations of our senses.
\end{quotation}
However, this concept was and presumably is still rejected by most physicists. {\em Hans Reichenbach}, for example, comments it as follows [3]:

\begin{quotation}
Newton's argument was severely criticized by Ernst Mach, who showed that it involved a serious {\em non sequitur}. Newton noted quite correctly that the variations in the shape of the surface of the water are not connected with the rotation of the water relative to the sides of the bucket. But he concluded that the deformations of the surface must therefore be attributed to a rotation relative to {\em absolute space}. However, this conclusion does not follow from the experimental data and Newton's other assumptions, for there are in fact two alternative ways of interpreting those data: the change in the shape of the water's surface is a consequence of either of a rotation relative to absolute space or of a rotation relative to {\em some system of bodies different from the bucket}.
\end{quotation}

As a matter of fact, intuition seems indeed to suggest that if uniform translation is a relative phenomenon then uniform rotation should be a relative process either. This expectation, however, is not justified at all, because there is a very significant difference between uniform translation and uniform rotation even at the level of pure kinematics. The relativity of uniform translation is related to the existence of a velocity addition law in both Newtonian and relativistic physics, which endows uniform translations with a Lie group structure. However, there is no analogous "angular velocity addition law" for rotations around axes intersecting at some common point because two consecutive uniform rotations in general lead to nonuniform rotation. As a result, uniform rotations around different axes do not form a Lie group, even though time-independent translations and rotations both have a Lie group structure. This fundamental difference between uniform rotations and uniform translations makes ideas based on the relativity of rotation untenable. 

In the 20th century, planetary orbits are interpreted as geodesics in space-time deformed by the huge mass of the Sun. To calculate them one is faced with the problem to choose the coordinate system best suited to the physical situation under study. In sharp difference with Newtonian treatment, there seems to be no place for choosing an inertial reference frame too in addition to the coordinate system because in general relativity no global inertial frames exist. In spite of these fundamental differences results of general relativistic calculations turn out very similar to those of Newton's theory of gravity, though they correct and implement them in several important respects.

The subject of the present note concerns this difference between Newtonian and general relativistic treatment of planetary motion and can be summarized in the following statement which justifies the established practice: {\em In general relativity the choice of inertial reference frame is replaced by the requirement that the solution be asymptotically Minkowski} up to a time-independent transformation of space coordinates. The Schwarzschild and the Kerr solutions obviously obey this criterion. Notice that it refers to coordinates rather than the geometric property of asymptotic flatness which is supposed to be the case.

These solutions can of course be rewritten in rotating coordinates --- there are problems that are easier to solve in a rotating coordinate system both in Newtonian gravity and general relativity. For example, when a point  mass revolves uniformly in a circular orbit around a spherically symmetric star of mass $M$, we obtain in rotating Schwarzschild coordinates
\begin{equation*}
\begin{split}
ds^2 = &\left [1-\frac{r_g}{r} - \left (\frac{r\omega}{c}\right )^2
\cdot\sin^2\vartheta\right ]c^2dt^2 - \\
& -2\omega r^2\sin^2\vartheta\:dtd\varphi - \frac{dr^2}{1-r_g/r} -
r^2(d\vartheta^2 + \sin^2\vartheta\:d\varphi^2),
\end{split}
\end{equation*}
the following answer for the relation between the radius $r$ of the rotation and its angular velocity $\omega$: 

\begin{equation*}
\Gamma_{tt}^r=\frac{r-r_g}{r}\left (\frac{MG}{r^2}-r\omega^2\right )=0.
\end{equation*}

In the language of Newtonian physics, formally (because $r$ is not a distance, but a coordinate), this condition exactly corresponds to the equality of gravitational attraction and centrifugal force.  However, in the theory of general relativity, the criterion of non-rotation cannot consist in the absence of the centrifugal force (a concept which is meaningless outside local frames), but rather in the smooth, continuous, singularity-free transition of the metric around the star into the asymptotic Minkowski metric. The rotating coordinate system does not meet this condition because it has a characteristic coordinate singularity: The component $g_{00}$ changes sign at finite $r$ while $g_{t\varphi}\longrightarrow\infty$ as $r\longrightarrow\infty$. It is for this reason that the line element above is related to a nonrotating object in rotating coordinates rather than the other way around.

In general relativity, considering an isolated star is an accepted idealization [4], although strictly speaking such objects don’t exist in the Universe. According to this assumption, the space-time of a solitary star ${\cal S}_1$ can be considered asymptotically flat, because the influence of other stars and the cosmological curvature of space-time can be disregarded with high accuracy.

Let us now consider another star ${\cal S}_2$ in this asymptotic domain. This star will usually move with some constant velocity $\vec v$ in the asymptotically Minkowski coordinates ${\cal K}_1$ of the star ${\cal S}_1$. The asymptotic Minkowski coordinates ${\cal K}_2$ belonging to ${\cal S}_2$ will therefore differ from the asymptotic coordinates of ${\cal K}_1$ in a Lorentz transformation corresponding to the stellar velocity. In the region between the two stars, distant from both, the homeomorphism between the two maps ${\cal K}_1$ and ${\cal K}_2$ will, therefore, be the Lorentz transformation corresponding to the speed of ${\cal S}_2$\footnote{Relation of this kind of description to the Standard Model of Cosmology is unclear because the Einstein-Straus problem [5] has yet no universally accepted solution. The description of the Universe as the multitude of overlapping coordinate patches of individual stars is in a sense dual to the Standard Model which is based on the notion of the cosmological fluid of smoothed-out matter. From this point of view the homeomorphisms we are speaking about may be slightly different from Lorentz-transformation but we hope that this quantitative uncertainty does not invalidate our argumentation.}. The roles of the two stars are obviously interchangeable, so their translational motion is {\em relative}\footnote{If ${\cal S}_2$ has an angular momentum, this homeomorphism must also contain a spatial rotation, which is determined by the direction of the $z$-axis of ${\cal K}_2$ relative to the axes of ${\cal K}_1$.}.

Their rotation is, however, absolute. In the common domain, the two overlapping coordinate systems are Minkowski, so they rotate neither by themselves nor relative to each other. They are the asymptotics of the coordinate systems in the near stellar regions which on their part are continuations of the inner solution of the Einstein equations. The metric of the internal coordinate system fixes the energy-momentum tensor and from the components of this tensor, we can calculate the motion of the material the star is made of relative to the overlapping common nonrotating coordinate system described at the beginning of this paragraph. If, for example, the outer solution is Kerr, then the star's material is generally rotating in an absolute sense.

If this argumentation is correct, then Newton's problem ceases to be a problem. To illustrate this, consider the calculation and measurement of the geodetic precession of a gyroscope orbiting ${\cal S}_1$ as the 21st-century version of the bucket experiment. The calculation is performed in the Schwarzschild or Kerr coordinates of ${\cal S}_1$. But since these coordinate systems only exist in our imagination, we relate the gyroscope's orientation to the direction of the very distant {\em guide star} ${\cal S}_2$. According to experience, the precession rate we measure corresponds to the calculation within experimental error. Why?

Because the coordinate system ${\cal K}_1$ used for the calculation is actually not independent from that of the {\em guide star}, since it is part of an overlapping {\em singularity-free} coordinate system consisting of two maps (coordinate patches) that contains both objects and does not rotate. As for the requirement of regularity, it is made possible by the fact that the general theory of relativity is a {\em field theory}, in which this is a meaningful prescription for the field quantity $g_{ij}(x)$. In Newtonian physics, which is a particle theory, no such possibility exists, and it is this fundamental difference which makes connecting ${\cal S}_1$ with the very distant ${\cal S}_2$ a much more difficult task.

In the solution of electrodynamic problems, the lack of singularity in regions free of charges is a natural requirement because Maxwell's equations describe the dynamics of a special type of continuous {\em substance}, the electromagnetic field. Today in general relativity no such background is felt behind $g_{ij}(x)$. Moreover, the equivalence of coordinate systems seems even to prohibit this kind of interpretation of the metric tensor.

Coordinate systems are, however,  equivalent to each other only in that either they all satisfy Einstein's equations, or none of them do. In other respects they may be distinguished from each other and in calculations this possibility is systematically exploited. For example, we can require that the coordinate system reflect the symmetries of the physical object whose space-time is being examined. Since coordinate systems, which are equivalent from the point of view of the Einstein equations, may differ from each other in coordinate singularities, we are allowed, if necessary, to require that they do not contain certain types of coordinate singularities.

If we perceived vacuum as a continuous substance rather than pure void or emptiness then the absence of singularity in regions of space without masses would become as natural property of the metric as the continuity of ${\vec E}(x)$ and ${\vec B}( x)$ in charge free regions.

In 1920, in his now little-known Leyden lecture [6], Einstein argued that general relativity actually requires the rehabilitation of the ether hypothesis, which would play the role of such a substance\footnote{{\em L. Kostro} published several articles and a book about the evolution of Einstein's conception of the ether. In [7] one can find references to these works.}:

\begin{quotation}
[T]here is a weighty argument to be adduced in favour of the ether hypothesis. To deny the ether is ultimately to assume that empty space has no physical qualities whatever. The fundamental facts of mechanics do not harmonize with this view. For the mechanical behaviour of a corporeal system hovering freely in empty space depends not only on relative positions (distances) and relative velocities but also on its state of rotation, which physically may be taken as a characteristic not appertaining to the system itself. In order to be able to look upon the rotation of the system, at least formally, as something real, Newton objectiveses space. Since he classes his absolute space together with real things, for him rotation relative to an absolute space is also something real. Newton might no less well have called his absolute space "ether”; what is essential is merely that besides observable objects, another thing, which is not perceptible, must be looked upon as real, to enable acceleration or rotation to be looked upon as something real.
\end{quotation}

As a consequence of the absence of privileged coordinate systems (rest system, for example), this ether, whose existence is compatible with both special and general relativity, must differ from the pre-relativistic concept of ether in that no state of motion or rest can be attributed to it [5]:

\begin{quotation}
Not every extended conformation in the four-dimensional world can be regarded as composed of world threads. The special theory of relativity forbids us to assume the ether to consist of particles observable through time, but the hypothesis of ether in itself is not in conflict with the special theory of relativity. Only we must be on our guard against ascribing a state of motion to the ether.
\end{quotation}

Developments in physics in the last hundred years support Einstein's concept of the ether: The vacuum cannot be mere emptiness, because according to quantum field theory, it has {\em physical} properties. In Einstein's equation, the term containing the cosmological constant can be interpreted as the energy-momentum tensor of the vacuum [8]:

\begin{equation*}
T^{(vac)}_{ij}(x) = \frac{c^4}{8\pi G}\Lambda g_{ij}(x).
\end{equation*}

This conception, which today is only a plausible hypothesis, might be the indication of a connection between the physical properties of the vacuum and the metric tensor. Since $\nabla_ig_{jk}=0$ no 
four-vector can be formed from $T^{(vac)}_{ij}$ by differentiation and so its existence is compatible with the requirement that no state of motion be attributable to the ether of general relativity.

\medskip\medskip

{\large\bf References}

\medskip

[1] Newton, I.: Principia, Book 1, Scholium, English translation by I. Bernard Cohen and Anne Whitman, University of California Press (2016)

[2] \href{https://en.wikipedia.org/wiki/Lorentz\_ether\_theory}{https://en.wikipedia.org/wiki/Lorentz\_ether\_theory}

[3] Reichenbach, H.: Problems in the Logic of Scientific Explanation, p. 209. New York: Harcourt, Brace and World (1961)

[4] Wald, R. M.: General Relativity, ch. 11. The University of Chicago Press Chicago and London (1984)

[5] Einstein, A. and Straus E. G., Rev. Mod. Phys. 17, 120:

\href{https://doi.org/10.1103/RevModPhys.17.120}{https://doi.org/10.1103/RevModPhys.17.120}

[6] Einstein, A.: Äther und relativitätstheorie. Springer (1920). English translation: \href{https://einsteinpapers.press.princeton.edu/vol7-trans/17}{https://einsteinpapers.press.princeton.edu/vol7-trans/17}

[7] Kostro, L.: \href{https://link.springer.com/chapter/10.1007/978-1-4615-2560-8\_22}{https://link.springer.com/chapter/10.1007/978-1-4615-2560-8\_22}

[8] Weinberg, S: Cosmology, Oxford University Press (2008)

\end{document}